\documentclass[aps,twocolumn,showpacs,amsmath,amssymb,pra,superscriptaddress,floatfix]{revtex4-1}
\usepackage{braket}
\usepackage{amsmath}
\usepackage{amssymb}
\usepackage{mathtools}
\usepackage{graphicx}
\usepackage{dcolumn}
\usepackage{bm}
\usepackage{multirow}
\usepackage{leftidx}
\usepackage{color}
\usepackage{float}

\usepackage{amsfonts}
\usepackage{eufrak}
\usepackage[german,english]{babel}

\begin{document}
\title{Comment on ``Matter-Wave Interferometry of a Levitated Thermal Nano-Oscillator
Induced and Probed by a Spin"}

\author{F. Robicheaux}
\email{robichf@purdue.edu}
\affiliation{Department of Physics and Astronomy, Purdue University, West Lafayette,
Indiana 47907, USA}

 \date{\today}

\maketitle

References~\cite{scala2013matter,wan2016tolerance} propose an experiment
involving the center of mass (COM) position of a nanodiamond and the spin of its NV center
to demonstrate ``the interference between spatially separated states of the center of
mass of a mesoscopic harmonic oscillator ... by coupling it to a spin and performing solely spin
manipulations". The nanodiamond is held in a harmonic potential and
also feels the force from gravity. The idea is to use a spatially varying magnetic field
$\vec{B}=B_0(-x\hat{x}-y\hat{y}+2z\hat{z})$ to couple the spin and COM degrees of freedom.
In this comment, we show that the results of the proposed experiment
do not result from an entanglement between the spin and the COM
position, and, hence, a measurement on the spin part of the
wave function can not give information about the COM separation
of the $\pm 1$ states.

The conceptual problem is that the nanodiamond is not oscillating about the $z=0$ point
of the harmonic potential but about the shifted
position, $-z_0$, due to gravity. The spatial shift is of order $10^{-9}$~m.
To compare, the {\it separation} of the $|\pm 1>$ states due to the difference in the
force on them from the spatially dependent magnetic field
is of order $10^{-13}$ m.
At the shifted position, $\vec{B}$ is {\it nonzero} which leads to a trivial phase
accumulation between the $+1$ and $-1$ states. It is exactly this phase difference
which leads to Eq.~(11,10) in Refs.~\cite{scala2013matter,wan2016tolerance}. This
phase difference can be simply cancelled using a small magnetic field offset and
is not the result of entanglement between the position and spin operators; we note
that the constant term in the magnetic field, Eq.~(1) of
Refs.~\cite{scala2013matter,wan2016tolerance},
is specifically not included in their discussion although it leads to exactly
the same kind of phase difference.

The Hamiltonian, Eq.~(4) of Refs.~\cite{scala2013matter,wan2016tolerance}, is
rewritten using the {\it shifted} COM coordinate $\tilde{z}\equiv z+z_0$ as
\begin{equation}
H=DS^2_z + \hbar\omega_z\tilde{c}^\dagger\tilde{c} - 2\lambda S_z (\tilde{c}^\dagger +\tilde{c})
+\sqrt{\frac{2m\omega_z}{\hbar}}z_0 2\lambda S_z - E_s
\end{equation}
where $\tilde{z}=\sqrt{\hbar/(2m\omega_z)}(\tilde{c}^\dagger + \tilde{c})$, $S_z$ is in units of $\hbar$,
$DS^2_z$ is from the NV center anisotropic spin interaction, $\tilde{c},\tilde{c}^\dagger$
are the lowering and raising operators in the shifted coordinate system $\tilde{z}$,
$\omega_z$ is the COM oscillation frequency, $\lambda = B_0 g_{NV}\mu_B\sqrt{\hbar /(2m\omega_z)}$,
$B_0$ is from the spatially varying magnetic field above, $g_{NV}$
is the Land\'e factor of the NV center, $\mu_B$ is the Bohr magneton,
$z_0 \equiv g\cos (\theta )/\omega_z^2$, $\theta$ is the angle between the $z$-axis and vertical,
and the constant $E_s = (1/2)m\omega_z^2z_0^2$.
For the discussion below, the first three terms will be grouped into $H_1$, the
fourth term will be defined as $H_2$, and the fifth term is a constant and, thus, can be dropped.

The time propagation of the wave function can be found exactly.
Most importantly, the operators $H_1$ and $H_2$ commute. This means the wave function
propagation can be written exactly as
\begin{equation}
\Psi (t) = \exp (-iH_2 t/\hbar )\exp (-iH_1 t /\hbar ) \Psi (0)
\end{equation}
where $\Psi (0) =\psi_0 (\tilde{z})(|+1\rangle + |-1\rangle )/\sqrt{2}$ is an initial spatial
function times the symmetric combination of spins $+1$ and $-1$.
One can use the methods in Refs.~\cite{scala2013matter,wan2016tolerance} to solve for
the time dependent wave function or one can decompose $\Psi (0)$ into
the eigenstates of the $H_1$ operator. After an integer $N$ periods, $t=2\pi N/\omega_z$, the
\begin{equation}\label{eqsubmain}
e^{-iH_1 t /\hbar } \Psi (0) = e^{i N\eta }\Psi(0)
\end{equation}
where $\eta =8\pi\lambda^2/(\hbar\omega_z)^2-2\pi D /(\hbar\omega_z)$.
Thus, the part of the Hamiltonian that contains both the $S_z$ and the $\tilde{z}$ operators,
which is the only part of $H$ that can entangle the spin and
COM degrees of freedom, gives {\it no effect} on the wave function after an
integer number of periods.

However, the term from $H_2=\sqrt{2m\omega_z/\hbar}\;z_0 2\lambda S_z$ gives
\begin{equation}\label{eqmain}
e^{-iH_2 t/\hbar }\Psi (0) = e^{-iN\phi /2}\psi_0 (z)\frac{|+1\rangle + e^{iN\phi}|-1\rangle }{\sqrt{2}}
\end{equation}
after $N$ periods, where $\phi = 8\pi\lambda z_0\sqrt{2m\omega_z/\hbar}/(\hbar\omega_z)$. Evaluating
$\phi$ and  $\Delta\phi_{\rm grav}$ in Eq.~(10) of Ref.~\cite{scala2013matter} or
Eq.~(9) of Ref.~\cite{wan2016tolerance}, one can show that $\phi = \Delta\phi_{\rm grav}$. Thus,
the main result of Ref.~\cite{scala2013matter}, Eq.~(9), or the equivalent
Eq.~(8) of Ref.~\cite{wan2016tolerance}, is exactly obtained in Eq.~(\ref{eqmain}).
Since $H_2$ is proportional to $S_z$, has no dependence
on $\tilde{z}$, and commutes with $H_1$,
it {\it can not} contain information about the spatial degrees
of freedom. Also, the $H_2$ can be exactly cancelled by
a uniform magnetic field which means the measurement
is not probing the {\it separation} of the $\pm 1$ states.
Thus, the proposed
measurement would not give information about
``spatially separated states of the center
of mass of a mesoscopic harmonic oscillator".

\end{document}